\documentstyle[12pt]{article}
\newlength{\extraspace}
\newlength{\extraspaces}
\newcommand{\be}{\begin{equation}
\addtolength{\abovedisplayskip}{\extraspaces}
\addtolength{\belowdisplayskip}{\extraspaces}
\addtolength{\abovedisplayshortskip}{\extraspace}
\addtolength{\belowdisplayshortskip}{\extraspace}}
\newcommand{\ee}{\end{equation}}

\newcommand{\bq}{\begin{eqnarray}
\addtolength{\abovedisplayskip}{\extraspaces}
\addtolength{\belowdisplayskip}{\extraspaces}
\addtolength{\abovedisplayshortskip}{\extraspace}
\addtolength{\belowdisplayshortskip}{\extraspace}}
\newcommand{\eq}{\end{eqnarray}}
\newcommand{\ra}{\rightarrow}

\newcommand{\newsection}[1]{
\vspace{15mm}
\pagebreak[3]
\addtocounter{section}{1}
\setcounter{equation}{0}
\setcounter{subsection}{0}
\setcounter{footnote}{0}
\begin{flushleft}
{\large\bf \thesection. #1}
\end{flushleft}
\nopagebreak
\medskip
\nopagebreak}

\begin{document}

\addtolength{\baselineskip}{.8mm}

\begin{flushright}
{\sc PUPT}-1415\\
hep-th@xxx/9311164 \\
 November  1993
\end{flushright}
\vspace{.3cm}

\begin{center}
{\Large  Hot  gauge theories  and $Z_{N}$ phases.}\\
\vspace{0.4in}
{\large Ian I. Kogan}
\footnote{ On  leave of absence
from ITEP,
 B.Cheremyshkinskaya 25,  Moscow, 117259,     Russia.} \\
\vspace{0.2in}
{\it  Physics Department, Princeton  University \\
 Princeton, NJ 08544
 USA} \\
\vspace{0.7in}
{\sc  Abstract} \\
\end{center}

\noindent

In this paper the  several aspects of the $Z_{N}$ symmetry
 in gauge theories at  high temperatures are discussed.
 The metastable $Z_{N}$ bubbles in the
$SU(N)$ gauge theories with fermions may have, generically,  unacceptable
thermodynamic behavior. Their free energy $F \propto T^4$ with a positive
proportionality constant. This leads not only to negative pressure but also
to negative specific heat and, more seriously, to negative entropy.
We argue that although such domains are important in the Euclidean theory,
they cannot be interpreted as physical domains in Minkowski space.
 The related problem is connected  with the analysis of the high-temperature
 limit of the confining phase. Using the two-dimensional QCD with
 adjoint fermions  as a toy model we shall demonstrate that in the
 light fermion limit in this theory there is no breaking of the
 $Z_{N}$ symmetry in the high-temperature limit and thus  there are no
   $Z_{N}$ bubbles.

\vfill

\newpage
\newsection{Introduction.}

\renewcommand{\footnotesize}{\small}

\noindent

The ${{\rm Z}_N}$ symmetry of pure Yang-Mills theories
   plays an important
role in the study of their thermal properties \cite{zn}.
  It is known both from  perturbative studies \cite{gpy,nw1}  and from
lattice simulations  \cite{lattice}  that at high temperature the
${{\rm Z}_N}$ symmetry is spontaneously broken and the the Euclidean
theory has $N$ degenerate vacua distinguished by different
 vacuum expectation values of Polyakov line  \cite{zn}
\bq
	<L> =  <\frac{1}{N}{\rm tr}~P~e^{i\int_0^\beta A_0 d\tau}>
\label{L}
\eq
 In the presence of
 any matter  which
transforms as the fundamental representation of $SU(N)$
  (for example quarks in QCD or quarks, leptons and
 Higgs particles in  the  electroweak theory)  this
${{\rm Z}_N}$ symmetry is no longer present and all but one of these
$N$ degenerate vacua become
either metastable or unstable (depending on $N$ and the number of
  flavors).
Following
work on the computation of the interface
tension between phases of different ${{\rm Z}_N}$ ``vacua'' in pure
gluonic theories \cite{bhatt1}
it has recently been
argued \cite{bubbles} that these metastable vacua
may lead to interesting cosmological consequences.

 However  it  has  been found
in    \cite{bksw} that
   ${{\rm Z}_N}$ domains, generically,  have unacceptable
thermodynamic behavior, for example,
  their free energy $F \propto T^4$ with a positive
proportionality constant. This leads not only to negative pressure but also
to negative specific heat and, more seriously, to negative entropy
  which   means that something is definitely wrong in our understanding
 of the  hot gauge theories structure.
   Using the abovementioned thermodynamical arguments
  one must conclude that the ${{\rm Z}_N}$ -bubbles, i.e.
 different  ${{\rm Z}_N}$ phases  coexisiting in space, do not exist
 in  models with metastable vacua.
 However the thermodynamical arguments do not forbid the
 existence of the different degenerate ${{\rm Z}_N}$ phases
 in the theories with unbroken ${{\rm Z}_N}$ symmetry.
  It is completely unclear what will be wrong with the theory
 if one adds some matter in fundamental representation ?
 How this matter can destroy the  existence of  (meta)stable states ?
 In some sense  the situation would be much more clear
 if one can assume that these phases cannot coexist in the
 space even in the case of unbroken  ${{\rm Z}_N}$ symmetry.
  The problem of the $Z_{N}$ bubbles have been  recently
  discussed in \cite{gp}.

  Recently  Smilga argued in  a paper  \cite{smilga} that
 this  point of view  may be  indeed correct
  and that  "different
  ${{\rm Z}_N}$ thermal vacua of hot pure Yang-Mills
 theory distinguished in the standard approach by
 different values of Polyakov loop average correspond
 actually to one and the same physical state".
 He presented different arguments supporting this
 statement including the possible role of the
 infrared divergences in the calculation  \cite{bhatt1}
  of the   surface tension of the walls separating different
 ${{\rm Z}_N}$ phases as well as the difference between
 strong coupling lattice $SU(N)$ gauge  theories,
 where  ${{\rm Z}_N}$-bubbles exist indeed  \cite{zn},
 and the weak coupling continuum limit.

 Besides these general  arguments he considered  an
  interesting example of the $1+1$ dimensional hot  QED - the
 Schwinger model - where the similar problem
 appears. Instead of $Z_{N}  = \pi_{1}\bigl[SU(N)/Z_{N}\bigr]$
 different  vacuum states in a pure
 $SU(N)$ gauge theory one has  $Z = \pi_{1}[U(1)]$
 different states in the Schwinger model. However it   was
 shown  in \cite{smilga} that in this case there are no
   domain wall solutions with finite surface tension.

 In this paper we shall   consider another  $1+1$ dimensional
  model  which, contrary to the Schwinger model,  shares some
 common features with realistic $3+1$ dimensional gauge theories.
 This is the $1+1$ dimensional  $SU(N)$ gauge theory with
   Majorana fermions in the adjoint representation
 with the action
\bq
S_{adj} = \int d^{2}x Tr \big[
-\frac{1}{4g^{2}} F_{\mu\nu}F^{\mu\nu} +
 i\bar{\Psi}\gamma^{\mu}D_{\mu}\Psi + m\bar{\Psi}\Psi\big]
\label{adjaction}
\eq
  which obviously
 has  ${{\rm Z}_N}$ symmetry.  The light-cone quantisation
 of this  theory was considered in a
 large $N$ limit in  \cite{adj}. The spectrum consists of
 closed-string excitations. Contrary to the 't Hooft
 model  \cite{thooft}  with fermions in the fundamental
 representation of $SU(N)$  describing the open-string
 excitations with the only meson Regge trajectory, in this
 theory there is an infinite number of the closed-string
 Regge trajectories and the density of  particle states
 increases exponentially with energy
  \cite{kutasov}, \cite{bdk}
\bq
 n(m) \sim m^{-\alpha}\exp(\beta_{H} m)
\eq
 which means  that there is the
 Hagedorn temperature $T_{H} = \beta_{H}^{-1}$
   and the model undergoes a
   confinement - deconfinement
 transition at this temperature which must be the simplest
 analog of the real confinement-deconfinement transition
 in QCD. The numerical value of $\beta_{H}
 \approx (0.7 - 0.75) \sqrt{\pi/(g^{2}N)}$ in the large $N$
 limit was calculated in \cite{bdk}. The same  picture
 was obtained in a  recent paper \cite{dkb} where the
 Hagedorn spectrum was obtained for a $1+1$- dimensional
 QCD with adjoint scalar matter. The numerical value
 of inverse Hagedorn temperature in this case is
  $\beta_{H}
 \approx (0.65 - 0.7) \sqrt{\pi/(g^{2}N)}$.

  In a string theory  language the  Hagedorn transition
  occurs    due to the fact that
     some winding modes in the imaginary
 time direction  become tachyonic at high temperature
  \cite{vortices}, \cite{aw}) (see also \cite{ak}).
 In a  string description of the gauge tyheory this winding modes
  are generated  by  the Polyakov line operators
\bq
 L_{k}(\vec{x}) = \frac{1}{N} {\rm tr}~P~\exp
\big(i\int_0^{k\beta} A_0(\tau,\vec{x}) d\tau\big)
\label{polyakovl}
\eq
wrapping $k$ times around the imaginary time direction  $\tau$.
 In a very interesting paper \cite{pol} Polchinski
      studied the high-temperature limit of
 the confinig phase and calculated in large $N$ limit the
 mass of the tachyonic winding modes at high tempreatures.
    His method was used
 by Kutasov in \cite{kutasov} to study the stabilty
 of the confining phase in the two-dimensional QCD coupled
 to adjoint matter.  We would like to note that
  both  Polchinski and Kutasov analysis were based
  on the form of effective potential which leads to the
 existense of the $Z_{N}$ bubbles and these two problems
-  $Z_{N}$ bubbles and tachyonic winding modes seem to be
 ultimately related.

  In section 2 we shall  discuss the effective
 potential for Polyakov line and the thermodynamical
 properties of the ${{\rm Z}_N}$  phases. In section 3
 we  consider  the  Polchinski method and discuss
 its relation to the existense of the ${{\rm Z}_N}$
 phases. In section 4 the $1+1$ gauge theory
 with adjoint fermions will be considered. Using the
 results obtained in \cite{bvw}
 we shall demonstrate that  there are no
 coexisting $Z_{N}$ phases in this model by the same
 reasons which  have been  found by Smilga
 for Schwinger model \cite{smilga}. Moreover, there
 is no $Z_{N}$ symmetry breaking in this model in
 te light fermion limit $m << \sqrt{g^{2}N}$.
  In conclusion we shall discuss the obtained results
 and unsolved problems. In particular we shall consider
 the possibility that the metastable $Z_{N}$ phases
(if they do exist in the four-dimensional theories)
 may have interpretation as  states with inverse
 population, i.e. with negative temperatures.

\newsection{${{\rm Z}_N}$ domains in gauge theories.}

Let us briefly review the origin of the $Z_{N}$ structure
 in gauge theories.    It is most illuminating to begin
 with the Hamiltonian theory in $A_0=0$ gauge
and impose Gauss' Law $D_iE_i-g\psi^\dagger\psi=0$ as a constraint
on the Hilbert space of states.
The projection operator $P$ onto these states
is just the projection operator onto gauge invariant states:
\bq
P={{\int {\cal D} g ~U_g }\over {\int {\cal D} g }}\label{projection}
\eq
where the integral is over all gauge transformations $g(\vec r)$ with
$g(\vec r)\in SU(N)$ using the Haar measure ${\cal D} g$.
$U_g$ is the representation of the gauge
 transformation $g$ on the Hilbert space
of states.
The partition function at nonzero temperature $1/\beta$ is given by
\bq
Z={\rm Tr~}\left\{e^{-\beta H}P\right\}=\int {\cal D}g~{\rm Tr~}
	\left\{e^{-\beta H}U_g\right\}
\eq
If we compute this trace in a basis $\{\vert A_i, \xi\rangle \}$
where $\xi$ represents
an appropriate fermionic state then
\bq
	Z=\int {\cal D}A_i {\cal D}\xi\int {\cal D}g~
	\langle A_i, \xi\vert e^{-\beta H} U_g \vert A_i, \xi\rangle
\label{basistrace}
\eq
One can now proceed with the usual derivation of the
Euclidean functional integral
at nonzero temperature except to note that the presence of the factor
$U_g$ modifies the boundary conditions on both the gauge fields
and the fermions. Thus apart from an overall normalization
$$
	Z=\int {\cal D}g(\vec r)\int_{\rm B.C.}{\cal D} A_i(\vec r, \tau)
	{\cal D}\psi(\vec r, \tau) {\cal D}\psi^\dagger(\vec r, \tau)
	e^{-\int_0^\beta d\tau \int d^3r~\cal L_E}
$$
\bq
	{\rm B.C.}:~~A_i(\vec r, \beta)= g\left[ A_i(\vec r, 0)\right]~~~~~
	\psi(\vec r, \beta)=-g\left[\psi(\vec r, 0)\right]
\eq
where $L_E$ is the usual Euclidean Lagrangian for a gauge theory coupled to
fermions, $g[A] \equiv gAg^{-1}-\partial gg^{-1}$ and
$g[\psi]\equiv g\psi$. In other words we derive the usual functional
integral in $A_0=0$ gauge but where the periodic (antiperiodic) boundary
conditions are modified to be periodic (antiperiodic) up to an arbitrary
gauge transformation which we then integrate over.

It is of course possible to remove these strange boundary conditions
by performing a gauge transformation in the functional integral.
For  each value of the integrand $g$
we may introduce any gauge transformation $V(\vec r, \tau)$ with the property
that  $V(\vec r, 0)=I$ and $V(\vec r, \beta)=g^{-1}(\vec r)$.
This will force the introduction of a temporal gauge field
$A_0=\partial_0V~V^{-1}$. The integral over $g$ will then become an
integral over $A_0$ with the appropriate measure. In fact if we integrate
over all possible such $V$'s we recover the usual path integral over
all gauge fields $A_\mu(\vec r, \tau)$ with periodic boundary conditions
and over all fermionic fields with antiperiodic boundary conditions.

If we consider
 spatially constant gauge transformation $g\in {{\rm Z}_N}$ i.e.
$g=\exp(2\pi ik/N)\times I$ (where $k$
 is an integer and $I$ is the identity matrix). Then
$g[A_i]=A_i$ but $g[\psi]=\exp(2\pi ik/N)\psi$.
 Thus in the absence of fermions
there are $N$  degenerate  vacua.
When fermions are present, however, a value of $g\in {{\rm Z}_N}$
  corresponds
to a path integral in which the fermions have the ``twisted'' boundary
conditions $\psi(\beta)=-\exp(2\pi ik/N)\psi(0)$. One can say
 that this describes
   fermions having an imaginary chemical potential (for more
 detailed discussion see, for example  \cite{nw2}).

The main tool for analyzing the ${{\rm Z}_N}$ structure of
  gauge theories
 is the calculation of the effective potential in a constant,
background temporal gauge field $A_0$.
 For definiteness we   begin by considering
an four dimensional   $SU(N)$ gauge theory with $N_f$ flavors .
 of massless Dirac
fermions. Although at finite temperature it is not possible
to choose a gauge in which $A_0=0$ it is possible to choose
a gauge in which $A_0$ is independent of (Euclidean) time $\tau$
and in which it is diagonal. $A_0$ can then be written as
\bq
	\Theta  =\beta  A_0 =
	\pmatrix{\theta_{1}&~& ~&  ~\cr
	~&  .& ~& ~\cr~&~&.&~\cr  ~& ~& ~& \theta_{N}}
\label{Theta}
\eq
where for $SU(N)$ one has $\theta_{1} +\ldots +\theta_{N} = 0~
 (\rm mod~ 2\pi/\beta)$
 and there  are  $N-1$ independent $\theta_{i}$ - the number of
 independent diagonal generators
of $SU(N)$ (i.e. the elements of the Cartan subalgebra of the Lie
Algebra of $SU(N)$). In this gauge the Polyakov line is given by
\bq
L   = \frac{1}{N} \sum_{i=1}^{N} e^{i\theta_{i}}
\label{l}
\eq

The effective potential for $A_0$ has been calculated up to two
loops \cite{gpy,nw1,twoloop}.  Here we shall  consider only the
one loop result  \cite{gpy,nw1}.
For gluons the effective potential is
\bq
V_{G}(\theta_{1},\cdots \theta_{N})
 =  -  \frac{\pi^{2}T^{4}}{45}(N^{2}-1) + \nonumber \\
\frac{\pi^{2}T^{4} }{24} \sum_{i,k =1}^{N}
\left(\frac{\theta_{i}}{\pi} - \frac{\theta_{k}}{\pi}
\right)^{2}_{\rm mod~2}
\left[ 2 - \left(\frac{ \theta_{i}}{\pi} - \frac{\theta_{k}}{\pi}
\right)_{\rm mod~2}\right]^{2}
\label{glpot}
\eq
and for each fermion flavor in fundamental representation it is
\bq
V_{F}(\theta_{1},\cdots \theta_{N})
 =   \frac{2\pi^{2}T^{4}}{45} N
 - \frac{\pi^{2} T^{4} }{12}  \sum_{i=1}^{N}
\left\{1 - \left[\left(\frac{ \theta_{i}}{\pi} +1 \right)_{\rm mod~2}
 - 1 \right]^{2} \right\}^{2}
\label{ferpot}
\eq

 The values of the effective potentials at zero field
  $V_{G}(0,\ldots, 0) = -(\pi^{2}T^{4}/45) (N^{2}-1)$
 and  $V_{F}(0,\ldots, 0) = -(7/4)(\pi^{2}T^{4}/45) N$. It is easy to
 see that it is
  the free energy density of an ideal gas
of gluons and fermions  at a temperature $T$ which
  is equal to $-\pi^2T^4\kappa/90$,
where each bosonic degree of freedom contributes 1 to $\kappa$ and
each fermionic degree of freedom contributes $7/8$ to $\kappa$.
Thus the gauge fields contribute $2(N^2-1)$
and the fermions contribute $4(7/8)N_fN$ to $\kappa$ which reproduces
$V_{G}(0,\ldots, 0)$ and $V_{F}(0,\ldots, 0)$.

The gluon effective potential (\ref{glpot})  has $Z_{N}$ symmetry
\bq
\theta_{i} \ra \theta_{i} + \frac{2\pi k}{ N}
\label{zn}
\eq
where integer $k = 0, \ldots, N-1$ must be the same for all $\theta_{i}$.
Then $\sum \theta_{i} = 0
{}~(\rm mod 2\pi/\beta)$ and without fermions
 there are $N$ minima at $\theta_{i} = (2\pi/\beta N) k$ where
 $k = 0, \ldots, N-1$. The fermion potential (\ref{ferpot})
 obviously violates the $Z_{N}$ symmetry (\ref{zn}) and only
 $\theta_{i} = 0$ is the global minimum of the total effective potential.
 All other $Z_{N}$ vacuua become either local minima
 (metastable states)  or unstable states  depending
 on the number of fermion flavors $N_{f}$.

 Furthermore it will be sufficient
for us to consider the case in which
 $\Theta$ has the form
\bq
	\Theta = \frac{2\pi}{ N} q
	\pmatrix{1&~& ~& ~& ~\cr ~& 1& ~& ~& ~\cr
	~& ~& .& ~& ~\cr~&~&~&.&~\cr ~& ~& ~& ~& -(N-1)}
\label{Theta1}
\eq
  The total free
 energy as a function of $q$
  is the sum of a  gluon (\ref{glpot})  and $N_{f}$
 fermion (\ref{ferpot}) effective potentials and can  be written
as
\bq
        {F(q)\over {\pi^2T^4}}={4\over 3}(N-1)\left(V_G(q)+N_fV_F(q)\right)
        -{1\over{45}}\left((N^2-1)+{7\over 4}NN_f\right)
\label{F}
\eq
where the bosonic and fermionic contributions are given
 in terms of the function
\bq
	f(x)=(x_{\rm mod~1})^2(1-x_{\rm mod~1})^2
\eq
by
\bq
	V_G(q)=f(q),~~~~~~
	V_F(q)=	{1\over 16}{N\over{N-1}}-f({q\over N}+{1\over 2})
	-{1\over{N-1}}f({q\over N}+{1\over 2}-q)
\eq
Notice that the function $f$ is periodic with period 1. Thus
$V_G(q)$ is periodic with period 1 but $V_F(q)$ is periodic
with period $N$. Note also that a constant has been added
to $V_{F}$ so that it vanishes at $q=0$ which is the
perturbative vacuum of the theory. The last  term
 in (\ref{F}) is
  the free energy density of an ideal gas
of gluons and $N_{f}$ fermions  at a temperature $T$.

Now let us repeat  the  arguments   which were used in \cite{bksw}
 to demonstrate very serious  problems arising
 in the thermodynamical description of the metastable $Z_{N}$
 vacuua.
For example for  $N=3$  it is easy to show that the metastable
minimum at $q=1$
remains metastable for $N_f<18$ at which point it becomes unstable.
Notice, however that for $N_f>3$ the free energy density becomes
{\bf positive} in these metastable states. Since $F\propto T^4$ this
poses a very serious problem which we shall now discuss.

First note that the positivity of the free energy at nonzero values
of $q$ is entirely due to the fermions.
For integer   $q \leq N/2$ one gets
\bq
	{{F(q~{\rm integer}~\le N/2)}\over {\pi^2T^4}}~=~
	NN_f\left\{ {2\over 3}{q^2\over{N^2}}\left(
	1-{{2q^2}\over {N^2}}\right)-{7\over{180}}\right\}
	-{{N^2-1}\over {45}}
\eq
This will be positive provided
\bq
	N_f> \frac{N^2-1}{45~N} \left[{2\over 3}{q^2\over N^2}
	\left(1- {2q^2 \over N^2}\right)-{7\over 180}\right]^{-1}
\eq
For $N=3$ and $q=1$, for example, we find that $F$
is positive (and propotinal to
$T^4$) for $N_f>3$.  For even $N$ and $q = N/2$ one has positive
 $F(N/2)$ for $N_{f} > (N^{2}-1)/2N$. This point will be
 a  local minimum for $N_{f} < N$. Thus for $N=4$ and $N_{2} =
 2,~3$  the point $q=2$ is both a minimum of $F$ and has a
 positive value $F$ propotional to $T^4$.  In more details
 this  was discussed  in \cite{bksw}.

It follows from the above discussion that for a large class of
models with both gauge fields and fermions
we can write the
free energy  density as
\bq
	F=+\vert\gamma\vert T^4
\eq

Such a situation is {\bf impossible} for the metastable state of
a real physical system. To see this let us remember that the free
 energy of any physical system at temperature $T$  is defined as
\bq
F(T) = -T \ln \sum_{n} e^{-E_{n}/T}
\eq
Shifting  $E_{n}$ by a constant $C$   one  can  add  $C$ to $F(T)$,
 but the $T$ dependent part must be negative as one can see
  defining energy levels $E_{n}$ in a such way  that the ground state
 energy $E_{0} = 0$  and $F(T) = -T \ln (1 + \ldots) < 0$.  In our case
 we got positive $F(T)$ which leads to physically senseless
 thermodynamic quantities, namely the
 negative  entropy density
\bq
        S={{E-F}\over T} = - \frac{dF(T)}{dT} =-4\vert\gamma\vert T^3,
\eq
the negative  internal energy density
\bq
        E= F + TS = -3\vert\gamma\vert T^4,
\eq
 the  negative  pressure
\bq
	p=-\vert\gamma\vert T^4
\eq
 and the negative  specific heat
\bq
	c=-12\vert\gamma\vert T^3.
\eq
Such a metastable vacuum thus has not only a negative pressure
but also negative specific heat and worst of all a negative
entropy. It is clear that no physical systems with positive
 temperature  of this type can
exist. However one cannot exclude  the metastable states
 with inverse population (the well known examples are lasers).
 We shall briefly discuss such an intriguing  possibility in a  conclusion.

There {\bf are} interesting cases in which the free energy density at
the metastable minimum has the correct, negative sign for $F$.
One such example is given in Ref. \cite{bubbles} in which the
standard electroweak model is considered well above the $QCD$
phase transition point. In this case the
base free energy density of the leptons, the Higgs and the weak gauge
 bosons contribute
to the total free energy density and make it negative.  It is clear, however,
that if a subsystem of the full system (namely the quarks and
gluons) has the disease discussed above, namely a positive free energy density
which grows like $T^4$, then including the leptons, Higgs and weak gauge
bosons which couple weakly to it cannot save the situation.
This will be clarified below but if we imagine a system
whose total entropy is positive  but that
 some identifiable subsystem has negative entopy   then any
 statistical description of the full system fails since
the subsytem has no states available to it.
More discussion about thermodynamical properties of these
 states as well
  as    difficulties  arising
 in interpretation of  $Z_{N}$ domains in Minkowski space  can
 be found in \cite{bksw}.

Let us note that the real problems arise only when we are
 assuming that this metastable states appear in our space
 as  $Z_{N}$ bubbles. To get such bubbles it was usually
 assumed that one can use the effective potentials (\ref{glpot})
 and (\ref{ferpot}) not only for  constant $A_{0}$ for
 which the potentials  had been really found in \cite{gpy}
 and \cite{nw1}, but also for  spatially dependent
 $A_{0}(\vec{x})$. Let us note that $Z_{N}$ symmetry
 (neglecting the matter in a  fundamental representation) exists
 only for  coordinate independent part  $A_{0}$ and generally
 speaking  it is not necessary that potential for nonconstant
 modes  $A_{0}(\vec{x})$ is the same. Moreover, as it has
 been demonstarted by Smilga in Schwinger model \cite{smilga}
  and as we shall  demonstrate later in a $1+1$ QCD
 with adjoint matter, there are situations when   the total
 effective potential is the sum of two independent potentials
 for constant and  nonconstant modes. In this case the
  zero mode $A_{0}$ becomes quantum mechanical variable
 and no $Z_{N}$ bubbles exist in space - however the price
 for this is the unbroken $Z_{N}$ symmetry. Before we
 shall  start this discussion let us  consider
  how one can use  the effective potential (\ref{glpot})
   to  analyse   the high-temperature
 limit of the confining phase  following the ideas suggested
 by Polchinski in \cite{pol}.

\newsection{$Z_{N}$ phases and high-temperature
 limit of the confinig phase.}

Let us consider the two-point correlation
 function of Polyakov lines (\ref{polyakovl})
 at low temperature in a pure gluodynamics
\bq
\left< L_{k} (\vec{x}) L_{-k}(0)\right> \sim
\exp\left(-M_{k}(\beta)x\right),
{}~~~ x \ra\infty
\eq
The correlation function vanishes at infinity because of
 the confinement,
 so the expectation value of  Polyakov line is zero
  $\left<L_{k}\right>  = 0$
 and $Z_{N}$ symmetry is unbroken.  The usual interpretation of the
 Polyakov line is the world line of an  external source in fundamental
 (for $k= 1$) or in higher ($k > 1$) representations.  However there
  is a dual description when one can consider compact Eucledian time
 $\tau$ as a spatial coordinate and $ L_{k}$ is the creation operator
 of a winding state with a electric flux in the periodic direction
 and $M_{k}(\beta)$ is
 a temperature dependent mass of the winding state.
  In the deconfinement phase the winding modes become tachyonic
 and the theory makes a transition to a phase with  broken $Z_{N}$
 symmetry.

Polchinski in \cite{pol}
  used the high-temperature effetctive potential
 (\ref{glpot}) to calculate the mass of these tachyonic states.
  To find them he considered the effective action
\bq
S_{eff} = \int d^{3} x \
 \frac{1}{2g^{2}\beta} \sum_{i=1}^{N} (\vec{\nabla}\theta_{i})^{2} +
 \nonumber \\
\frac{1 }{24\pi^{2}\beta^{3}} \sum_{i,k =1}^{N}
\bigl(\theta_{i} - \theta_{k}\bigr)^{2}_{\rm mod~2\pi}
\left[ 2\pi - \bigl(\theta_{i} -
\theta_{k}\bigr)_{\rm mod~2\pi}\right]^{2}
\label{effaction}
\eq
The gradient term corresponds  to the square of the electric field
 $\vec{E}^{2}$ in the bare action and the second term  can
 be obtained from the  effective potential (\ref{glpot}) after
 omiting  $\theta$ independent first term.  In the
 large $N$ limit one can  introduce the normalized density
\bq
\rho(\theta, \vec{x}) = \frac{1}{N} \sum_{i}\delta\left(\theta -
 \theta_{i}(\vec{x})\right)
\label{density}
\eq
In the $N \ra \infty$ limit $Z_{N}$ symmetry is transformed into
 $U(1)$ symmetry $\theta \ra \theta + {\rm const}$.
 It is easy to see that $L_{k}(\vec{x})$ are the  Fourier coefficients
 of the density $ \rho(\theta, \vec{x})$ (see (\ref{l}))
\bq
\rho(\theta, \vec{x})  =
 \frac{1}{2\pi} \left[ 1 + \sum_{k = 1}^{\infty}
  L_{k}(\vec{x}) \exp(-ik\theta)
+  L_{-k}(\vec{x}) \exp(ik\theta) \right] \;\;\;\;\;\;\; \nonumber \\
 L_{k}(\vec{x}) = \int_{0}^{2\pi} \rho(\theta, \vec{x})
e^{ik\theta} d\theta  =
 \frac{1}{N} \sum_{i=1}^{N} \exp\left(ik\theta_{i}
(\vec{x})\right)  = \frac{1}{N}{\rm tr}P\exp
\big(i\int_0^{k\beta} A_0(\tau,\vec{x}) d\tau\big)
\eq
The action (\ref{effaction}) takes the form
\bq
S_{eff} = \frac{N}{2g^{2}\beta}
 \int d^{3} x \int_{0}^{2\pi} d\theta
 \frac{1}{\rho(\theta, \vec{x})}
(\partial_{\theta}^{-1}
\vec{\nabla} \rho(\theta, \vec{x}))^{2}
  + \nonumber \\
\frac{ N^{2}}{24\pi^{2}\beta^{3}} \int d^{3} x
 \int_{0}^{2\pi}  d\theta_{1}  \int_{0}^{2\pi}  d\theta_{2}
 \rho(\theta_{1}, \vec{x})\rho(\theta_{2}, \vec{x})
\bigl(\theta_{1} - \theta_{2}\bigr)
\bigl[ 2\pi - \bigl(\theta_{1} -
\theta_{2}\bigr)\bigr]^{2}
\label{rhoaction}
\eq
where $\partial_{\theta}^{-1}$ is defined as
 $\partial_{\theta}^{-1} \exp(ik\theta) = \exp(ik\theta)/ik$
 and in large $N$ limit we must keep $g^{2}N$ fixed. The minimum
 of the potential is when all the eigenvalues are equal which
 gives us the spectral density in high-temperature  phase
 with broken $Z_{N}$ (here is $U(1)$ in large $N$ limit)
 symmetry
\bq
\rho_{\rm broken}(\theta, \vec{x}) = \delta(\theta - \theta_{0})
\eq
 for some $\theta_{0}$. The symmetric confining phase is defined
 by the $U(1)$ invariant distribution
\bq
\rho_{\rm c}(\theta, \vec{x}) = \frac{1}{2\pi}
\eq
which is unstable.
 One can easily find that in quadratic approximation
in $L_{k}$   the action (\ref{rhoaction}) takes the form
\bq
S = N^{2} \sum_{k = -\infty}^{\infty} \int d^{3}x
\left[ \frac{1}{2g^{2}N\beta k^{2}} \vec{\nabla} L_{k}
\vec{\nabla} L_{-k}  +  \frac{1}{24 \pi^{2}\beta^{3}} V_{k}
 L_{k}L_{-k}\right]
\label{Laction}
\eq
 where
$V_{k}$ is the  Fourier transform  of the potential
$ V(\theta_{1} - \theta_{2}) =
\bigl(\theta_{1} - \theta_{2}\bigr)_{\rm mod~2\pi}^{2}
\bigl[ 2\pi - \bigl(\theta_{1} -
\theta_{2}\bigr)_{\rm mod~2\pi}\bigr]^{2}$
\bq
V_{k} = \frac{1}{2\pi} \int_{0}^{2\pi} d\theta e^{ik\theta} V(\theta)
= \frac{1}{2\pi} \int_{0}^{2\pi} d\theta e^{ik\theta} \theta^{2}
 (2\pi - \theta)^{2} =  - \frac{24}{k^{4}}
\eq
and  one gets the tachyonic winding modes with masses
\bq
 M_{k}^{2} = - 2g^{2}N/\pi^{2}\beta^{2}k^{2}
\label{mass}
\eq
  One can make
 an assumption that the same spectrum of the tachyon masses
 must be reproduced in the string theory (if any) describing
 QCD. Let us note that this spectrum is different from the spectrum
 obtained in the usual string theory \cite{vortices}
 $\alpha'M_{k}^{2} = - n/6 + \beta^{2}/4\pi^{2}\alpha'$ where
 $n$ is some effective constant proportional to the number of
 the world-sheet degrees of freedom ($24$ for critical bosonic
 string).  Using (\ref{mass}) Polchinski made a conclusion
 that for QCD string the effective number of  degrees of freedom  grows
  with  temperature as $n_{eff}(\beta) \sim g^{2}(\beta) N/\beta^{2}$
 and the main conclusion of his analysis is that a string
 theory describing QCD  in the  large $N$ limit  must have a
 number of world-sheet degres of freedom which diverges at short
 distances \cite{pol}.

Let us note that this analysis can be repeated even in the case
 when there are $N_{f}$ fermions in the fundamental representation.
Then using the (\ref{ferpot}) one can  see that the fermion
 contribution to the effective action (\ref{effaction}) will be
\bq
S_{f} = - \frac{N_{f}}{12\pi^{2}\beta^{3}} \int d^{3} x
\sum_{i = 1}^{N}\left[\pi^{2} -\left((\theta_{i} + \pi)_{\rm mod 2\pi} - \pi
\right)^{2}\right]^{2}
\eq
 This term corresponds to the linear in $\rho(\theta, \vec{x})$
 contribution to the effective action
 for the density $\rho$
\bq
 S_{f} = - \frac{N N_{f}}{12\pi^{2}\beta^{3}} \int d^{3} x \int_{-\pi}^{\pi}
 d\theta
\rho(\theta, \vec{x})(\theta^{2} - \pi^{2})^{2} =
  N N_{f} \sum_{k} \int d^{3}x
  \frac{(-)^{k}}{k^{4} \pi^{2}\beta^{3}} L_{k}(x)
\eq
Including this term into (\ref{Laction}) we get
\bq
S = N^{2} \sum_{k = -\infty}^{\infty} \int d^{3}x
\left[ \frac{1}{2g^{2}N\beta k^{2}} \vec{\nabla} L_{k}
\vec{\nabla} L_{-k}  -   \frac{1}{24 \pi^{2}\beta^{3} k^{4}}
 L_{k}L_{-k} + \frac{N_{f}}{N} \frac{(-)^{k}}{k^{4} \pi^{2}\beta^{3}}
 L_{k}\right]
\eq
and it is easy to see that linear terms do not change the values of the
 tachyon  masses (\ref{mass}), but shift the fields $L_{k}$
\bq
L_{k} \ra L_{k} + \frac{(-)^{k}}{2} \frac{N_{f}}{N}
\eq
For small $N_{f}/N$ this shift is small and the  quadratic approximation
 for  the total action (\ref{rhoaction}) is still valid. One cannot
 use this approximation when $N_{f} \approx N$, i.e. precisely when
 the suppression factor for nonplanar diagrams
 $N_{f}/N$  is of  order  one
  and nonplanar diagrams have  the same order of magnitude as the  planar
  ones.  In latter case  the connection between  large $N$ gauge theory
 and a string  theory
  is much more
 problematic.

Thus we see that effective high-temperature theory  gives us an important
 information about (possible) string  description of the low-temperature
 confinement phase. This information was based on the form of the
 effective potential which may lead in some situation to physically
  senseless results. So we must be very accurate in dealing with
 this potential and  for this reason let us study the simplest model
 where one can hope to have  confinement-deconfinement
 transition  - $1+1$ dimensional QCD with adjoint matter
 \cite{adj}, \cite{kutasov} - \cite{dkb}.

\newsection{Two-dimensional QCD coupled to adjoint matter at
 high temperatures}

Let us repeat following Kutasov \cite{kutasov} the analysis
 of the previous section in the case of the $1+1$-dimensional QCD
 interacting with Majorana fermions in the adjoint representation
  described by the  action
\bq
S_{adj} = \int dx \int_{0}^{\beta}  d\tau Tr \left[
 \frac{1}{2g^{2}} F_{\mu\nu}F^{\mu\nu} +
 i \bar{\Psi}\gamma^{\mu}D_{\mu}\Psi + m\bar{\Psi}\Psi\right]
\label{newadjaction}
\eq
defined on  a Eucledian space-time with periodic Eucledian
 time $\tau \sim \tau + \beta$ with periodic boundary conditions
 for gauge fields and antiperiodic for fermions.  One can again choose the
 gauge  when $A_{0}$ is diagonal and independent on  time $\tau$:
$~A_{0} = (1/\beta){\rm ~diag}(\theta_{1},\ldots \theta_{N})$.  Then the
 one-loop
 effective action for the $\theta_{i}$ takes the form
\bq
S_{eff} = \frac{1}{2g^{2}\beta} \int dx \sum_{i=1}^{N}
 (\frac{d\theta_{i}}{dx})^{2} + V(\theta_{1},\ldots \theta_{N})
\eq
where the effective potential is nothing but the determinant
 of the Dirac operator  in adjoint reperesentation
 $\gamma_{\mu}D^{adj}_{\mu}$  in an external $A_{0}$ field
 \bq
V(\theta_{1},\ldots \theta_{N}) =
 - \ln det \left[\gamma_{\mu}D^{adj}_{\mu}[A] + m\right] =
 - \frac{1}{2} \sum_{i,j = 1}^{N}
\ln det \left[\gamma_{\mu}D_{\mu}[\theta_{i}-\theta_{j}] + m\right]
\eq
Using the proper time representation of the fermion determinant
 one gets
\bq
V(\theta_{i} - \theta_{j}) =
\frac{\beta}{2\pi} \int dx \sum_{k=1}^{\infty}(-)^{k} \int_{0}^{\infty}
\frac{d\tau}{\tau^{2}} \exp(-\frac{k^{2}\beta^{2}}{4\tau} - m^{2}\tau)
\cos k(\theta_{i} - \theta_{j}) \nonumber \\
V(\theta_{i} - \theta_{j}) = \frac{1}{2\pi\beta} \int dx
[(\theta + \pi)_{\rm mod 2\pi} - \pi]^{2}, ~~~~~~~~~  m  = 0
\label{vtheta}
\eq
and in the high-temperature limit $\beta \ra 0$ one can neglect
 the mass term in the leading $1/\beta$ approximation. The sum
 over $k$ is the sum over winding of the particle trajectory
 around the compact imaginary time
  and in the high-temperature limit $\beta \ra 0$ one can neglect
 the mass term  \footnote{it is easy to see that
 the corrections will be by  order of $m\beta$}
 in  the leading $1/\beta$ approximation.
This representation  of the determinant  can be easily generalised
 to higher dimensions where  one must substitute $d\tau/\tau^{2}$ by
 $d\tau/\tau^{d+1}$ in the case of  $d+1$ dimensional theory. Also in the
 case of bosonic degrees of freedom the factor $(-)^{k}$ is missing.
 One can easily reproduce (\ref{glpot}) and (\ref{ferpot}) using
 the proper time representation of the boson and fermion determinants.

Thus for small $\beta$   the effective potential for
 {\bf constant} $\theta_{i}$ is known and assuming, as
 usual, that for slowly varying $\theta$ one can simply
 substitute $\theta_{i}$ by $\theta_{i}(x)$ one gets the effective
 action (see \cite{kutasov})
\bq
S_{eff} = \frac{1}{2g^{2}\beta} \int dx \sum_{i=1}^{N}
 (\frac{d\theta_{i}}{dx})^{2} +
\frac{2}{\pi\beta} \int dx \sum_{i,j=1}^{N}
\sum_{k=1}^{\infty} (-)^{k}\frac{1}{k^{2}}\cos k\left[\theta_{i}(x) -
\theta_{j}(x)\right]
\label{2potential}
\eq
Obviously this action has $Z_{N}$ symmetry $\theta_{i}
 \ra \theta_{i} + 2\pi m/N,~m=1,2,\ldots N$ and we get the same
 $Z_{N}$ bubbles as in four-dimensional case. One can also consider
 the density (\ref{density})  $\rho(\theta, x)$ and using the
 same  method as in \cite{pol} Kutasov obtained the effective
 action for $L_{k}(x)$ in quadratic approximation \footnote{
 To get the effective action in quadratic approximation one must simply
 repeat the same procedure as in a $3+1$ dimensional case. Let us
 note also that because we need Fourier transformed  effective
 potential  $V_{k}$  it is not necessary to  take  the  sum over
 $k$ in (\ref{2potential}) to get $V(\theta)$.}
\bq
S = N^{2} \sum_{k = -\infty}^{\infty} \int d x
\left[ \frac{1}{2g^{2}N\beta k^{2}} \frac{d L_{k}}{dx}
  \frac{d L_{-k}}{dx}  + (-)^{k}  \frac{2}{\pi\beta k^{2}}
 L_{k}L_{-k}\right]
\label{2action}
\eq
 and got the masses of the winding modes
\bq
M^{2}_{k}(\beta \ra 0) = \frac{4 g^{2} N}{\pi} (-)^{k}
\eq
where only the odd $k$ winding modes are tachyonic.

However in the massless limit the form of the effective
 potential (\ref{2potential}) which leads both to the existence
 of the $Z_{N}$  domain walls  (and in the case of the fermions in the
 fundamental representation they will be metastable bubbles) and
 gives the imaginary mass of the unstable winding states is wrong.
 As we shall demonstrate now in the massless theory there are no
 domain walls and one cannot make simple predictions about the
 condensation of the winding states - and all this because the
potential for the $x$ dependent modes of the fields $\theta_{i}$
 is not periodic contrary to the constant mode part which was
 really calculated in (\ref{vtheta}).   The two-dimensional
 fermion determinant on cylinder
 in arbitrary gauge field  was calculated in \cite{bvw} where
 it was shown that it is factorized into product of two
 independent terms.  The first one
 is the periodical potential but for constant modes $\theta_{i}$,
  not for the fields $\theta_{i}(x)$. The second term depends on
    $x$-dependent  fields $\theta_{i}(x)$, however there is no
  reason why this part must have $Z_{N}$ symmetry - and it
 has not. In result any  configuration with domain wall has an
 energy proportional to the one-dimensional system length $\int dx$ and
 these configurations are absolutely irrelevant.

 To make this statement more clear let us repeat the calculation
 of the  determinat of the two-dimensional  Dirac  operator
 and demonstrate the  factorization property  following the  paper
 \cite{bvw}. We  shall not consider here the case of  an arbitrary
 gauge field $A_{\mu}$, where $A_{\mu}$ is a Hermitian matrix,
 instead we shall consider interesting for us case when this
 matrix is diagonal (see (\ref{Theta}))
\bq
	  A^{\mu}(\tau, x) =
	\pmatrix{A^{\mu}_{1}(\tau, x)&~& ~&  ~\cr
	~&  .& ~& ~\cr~&~&.&~\cr  ~& ~& ~& A^{\mu}_{N}(\tau,x)}
\label{diagA}
\eq
Then the fermion action  from (\ref{newadjaction})
takes the form
\bq
S_{f} = \int dx \int_{0}^{\beta}  d\tau Tr \left\{
  \bar{\Psi}\left(i\gamma_{\mu}\partial^{\mu} + m\right)\Psi +
\bar{\Psi}\gamma_{\mu}[A^{\mu}\Psi] \right\} = \nonumber \\
\int dx \int_{0}^{\beta}  d\tau \sum_{i,j}
  \bar{\Psi}_{ij}\left(i\gamma_{\mu}\partial^{\mu} +
 \gamma_{\mu}[A^{\mu}_{i} - A^{\mu}_{j}] + m\right)\Psi_{ij}
\eq
and we see that the total   determinant for the adjoint fermions
\bq
det (\gamma_{\mu}D_{adj}^{\mu}(A) + m) =
\prod_{i,j}det (\gamma_{\mu}[i\partial^{\mu} + A^{\mu}_{ij}] + m)
\eq
  is the product of abelian determinants of fermions interacting
 with abelian fields  $A^{\mu}_{ij} = A^{\mu}_{i} - A^{\mu}_{j}$.

As we discussed before in a high-temperature limit one can omit
 the mass term and what we are trying to calculate now is the
 determinant of the Dirac operator for the massless fermion
  in an  abelian gaauge field $A^{\mu}$
  on the cylinder $S^{1}\times R^{1}$
 or on the torus  $S^{1}\times S^{1}$ if the space is a circle too.
 Let us consider the  Hodge decomposition of the vector potential
(in the sector with zero topological charge)
\bq
A^{\mu} = \frac{1}{R_{\mu}} \theta^{\mu}  + \epsilon^{\mu\nu}
\partial_{\nu}\chi + \partial^{\mu}\xi
\eq
where the last term is the pure gauge and can be neglected.
The $R_{\mu}$ factors are the radii of two $S_{1}$ and
 $R_{\tau} = \beta$. In the
 case of the noncompact space, i.e. when we are considering
 the cylinder   $S^{1}\times R^{1}$ one can put  $R_{x} \ra \infty$.
  Constant
 modes $\theta^{\mu}$ are the angle variables with periodicity
 $2\pi$  and $\chi$ is a coordinate dependent mode.
Substituting the Hodge decomposition into Dirac operator
 and using the well known property of two-dimensional
 $\gamma$ matrices $\gamma_{\mu}\epsilon^{\mu\nu} =
i\gamma^{\nu}\gamma_{5}$
  one can get
\bq
i\gamma_{\mu}(\partial^{\mu} - iA^{\mu}) =
i\gamma_{\mu}\left(\partial^{\mu} - \frac{i}{R_{\mu}} \theta^{\mu}
-i\epsilon^{\mu\nu}\partial_{\nu}\chi \right) =
i e^{\gamma_{5}\chi}\gamma_{\mu}\left(\partial^{\mu} -
\frac{i}{R_{\mu}} \theta^{\mu} \right)e^{\gamma_{5}\chi}
\eq
Now let us consider a family of operators
\bq
{\cal D}_{\tau} =
i e^{\tau\gamma_{5}\chi}\gamma_{\mu}\left(\partial^{\mu} -
\frac{i}{R_{\mu}} \theta^{\mu} \right)e^{\tau\gamma_{5}\chi}
\label{Dtau}
\eq
We are looking for $D_{1}$ and $D_{0} =
\gamma_{\mu}\left(\partial^{\mu} -
\frac{i}{R_{\mu}} \theta^{\mu} \right)$
 is the operator in a constant field which determinant
 is periodic in $\theta^{\mu}$.

 A very elegant formula was obtained by Blau, Visser and Wipf
  in \cite{bvw} for
 a family of general first order elliptic self-adjoint operators
 depending on a parameter $\tau$ of the type
${\cal O}_{\tau} = \sqrt{g_{0}/g_{\tau}} \exp(\tau f^{\dagger})
{\cal O}_{0} \exp(\tau f)$ were $f(x)$  can be in general
 some matrix-valued function and $g_{\tau}$ is the parameter
 dependent metric on the manifold.  In our case (\ref{Dtau})
 we have $f(x) = \gamma_{5}\chi(x)$
 and flat metric independent on $\tau$.
 We shall present here the result for this  case only , more
 general expressions including general nonabelian field
 can be found in original paper \cite{bvw}.  For a family
 of   operators (\ref{Dtau}) one gets \footnote{
Here we neglect the possible contributions of the zero modes.
 If they are present the determinant itself is zero and one
 has the expression for $det'{\cal D}_{\tau}/det K_{\tau}$, where
  $ K_{\tau}$ is the matix of scalar products of zero modes.
 However if we  consider the sector with zero topologicaal charge
 there are no zero modes and one can simplify life a little}
\bq
\frac{d}{d\tau} \ln det{\cal D}_{\tau} = \frac{\tau}{\pi}
\int d^{2}x  \chi \partial^{2} \chi
\eq
 which can be easily integrated and finally we get
\bq
 det \gamma_{\mu}(\partial^{\mu} - iA^{\mu}) =
 det \gamma_{\mu}\left(\partial^{\mu} -
\frac{i}{R_{\mu}} \theta^{\mu} \right)
 \exp \left(\frac{1}{2\pi}
\int d^{2}x  \chi \partial^{2} \chi \right)
\label{det}
\eq
and we see that the contributions of  constant modes $\theta^{\mu}$
 and nonconstant modes $\chi(x)$ are factorizable ! Thus the
 effective potential is the sum of two terms - one is the
 effective potential for the constant field which we had calculated
 and which is periodic in $\theta$ with period $2\pi$. Another
 one is the potential for $x$-dependent part of the gauge field
 which {\bf is not} periodic at all - this is the ordinary
 Schwinger mass term proportional to $\int d^{2}x \chi \partial^{2} \chi
 = (1/4)\int d^{2}x \epsilon_{\mu\nu}F_{\mu\nu}(1/\partial^{2})
\epsilon_{\mu\nu}F_{\mu\nu}$.

Now let us consider the effective action  when only $A_{0}$
 component in  (\ref{diagA}) is nonzero and $A_{0}$ depends only
 on $x$. This is precisely the case which is relevant both for
 studying $Z_{N}$ bubbles and instability of the confining
 phase at high temperatures as we have discussed before.  Then  one can
 immediately write the effective action in which the
 constant modes $\theta_{i}$ are  completely independent from
 $x$ dependent fields $\tilde{A}_{i}$,  where $\tilde{}$
 means that we extracted the zero mode from $A_{i}(x)$.

\bq
S_{eff} =
 \frac{L}{2\pi\beta} \sum_{i,j}
[(\theta_{i} - \theta_{j}  + \pi)_{\rm mod~ 2\pi} - \pi]^{2} + \nonumber \\
  \frac{\beta}{2g^{2}} \int dx \sum_{i=1}^{N}
 (\frac{d A_{i}}{dx})^{2} +
\frac{\beta}{2\pi} \int dx \sum_{i,j=1}^{N}(\tilde{A}_{i}(x)
 - \tilde{A}_{j}(x))^{2}
\eq
where $L$ is the size of the one-dimensional system.
 The  $Z_{N}$ symmetry of this action  $\theta_{i}
 \ra \theta_{i} + 2\pi m/N,~m=1,2,\ldots N$  does not affect
 the $x$ dependent fields $\tilde{A}_{i}(x)$ at all.

It is clear now that for such potential any $Z_{N}$ bubbles
 will have infinite energy in the thermodynamical limit.
 To see it let us consider the field (see (\ref{Theta}))
\bq
        A_{0}(x) = \frac{2\pi}{\beta N} q(x)
        \pmatrix{1&~& ~& ~& ~\cr ~& 1& ~& ~& ~\cr
        ~& ~& .& ~& ~\cr~&~&~&.&~\cr ~& ~& ~& ~& -(N-1)}
\label{4.17}
\eq
where $q = 0$ and $q = 1$ corresponds to two different
 $Z_{N}$ vacuua. The effective action for one  interpolating
 field $q(x)$ is the following
\bq
S[q] = (N-1)\left\{  \frac{2\pi^{2}}{\beta g^{2}N} \int
 dx (\frac{dq}{dx})^{2}
 + \frac{2\pi}{\beta}   \int dx \tilde{q}^{2}(x)  +
\frac{2\pi L}{\beta}\left[\left(q_{0}+\frac{1}{2}\right)_{\rm
 mod~ 1} - \frac{1}{2}\right]^{2}\right\}
\eq
where $q(x) = q_{0} + \tilde{q}(x)$ and zero mode
$q_{0} = (1/L) \int_{0}^{L} dx q(x)$.  Let us consider the $Z_{N}$
 bubble, i.e. the region with $q(x) = 1$ with size $R$. The rest
 part  of the space, i.e. region with size $L - R$  has $q(x) = 0$
 and we  assumed that both $L$ and $R$ are large in comparison with
 domain wall width $1/\sqrt{N g^{2}}$, so one can neglect the
 contributions from the regions where $q(x)$ interpolates between $0$
 and $1$.   It is easy to see that in this
 case $q_{0} = R/L$ and $\tilde{q}(x) = q(x) - R/L$ and
   one must have  $R/L < 1/2$, in opposite case it is better to
  say that  there is  a bubble of $q = 0$ phase in the  space
 with $q = 1$. Then
  the second and third term in $S[q]$  contribute to the action
\bq
S(L,R) = (N-1)\frac{2\pi}{\beta} \left[ R(1-\frac{R}{L})^{2}
 + (L - R) (\frac{R}{L})^{2} + L (\frac{R}{L})^{2} \right] =
(N-1)\frac{2\pi}{\beta} R
\eq
and we see that the action is proportional to the bubble size $R$.
 To consider the single domain wall one must put  $R \sim L$
 sending the "anti"-wall  far apart, then in the thermodynamical  limit
 $L \ra \infty$ and single domain wall is infinitely heavy.

 Thus there are   no  $Z_{N}$ bubbles in this theory, moreover
 we  are not allowed to have any $Z_{N}$ vacua.
 The reason
 is very simple -  in this theory the $Z_{N}$ symmetry can not be
  broken spontaneously. The reason is obvious - due to the
  decoupling of constant modes from all other modes we  have
 to average over all $Z_{N}$ vacua and cannot restrict ourselves
 for only one - as it would be possible in the case without
 factorization.  Let us consider for example $SU(2)$ theory with
   the symmetry of the center $Z_{2}$. The Polyakov line
 in this case is \footnote{Here we are considering the
 simplest case with the winding number $k = 1$. The generalization
 for arbitrary $k$ is trivial}
\bq
L(x) = \frac{1}{2}[e^{i\pi q(x)} +  e^{-i\pi q(x)}] =
 \frac{1}{2}[e^{i\pi q_{0}}e^{i\pi \tilde{q}(x)} +
e^{-i\pi q_{0}}e^{-i\pi \tilde{q}(x)}]
\eq
and it is evident that $<L(x)> = 0$ because averaging over $q_{0}$
 is factorized and
\bq
<e^{\pm i\pi q_{0}}>_{q_{0}} = 0
\eq
 However the zero-mode $q_{0}$ factors are cancelled in
  the two-point correlation function
\bq
<L(x) L(0)> = \frac{1}{2} < \cos(\tilde{q}(x)-\tilde{q}(0))> =
 \frac{1}{2} e^{<\tilde{q}(x) \tilde{q}(0)>}
\eq
 which is non-zero and has finite limit at $x \ra \infty$.
 Thus we have $<L(x) L(0)> \ra 1/2$ when $x \ra \infty$
 and at the same time $<L(x)> = 0$. The breakdown of
 the clusterization property is connected with the
 factorization of the zero mode $q_{0}$. This mode
 is absolutely delocalised and one can not use naive
 rule $<A(x) B(0)> \ra <A(x)><B(0)>$ at $x \ra \infty$.

Because the  $Z_{N}$ symmetry is unbroken at high temperatures
 one can conclude that
  the analysis of the instability of the confining phase
 at high temperatures which was done in \cite{kutasov}
 is incomplete.  However we must remember that the
 factorization of the fermion determinant  (\ref{det})
 was obtained only for massless case. For massive fermions
 the effective action is
\bq
Tr \ln [\gamma_{\mu} D^{\mu} - m]
 = Tr \ln \gamma_{\mu} D^{\mu} - Tr \sum_{n=1}^{\infty}
 \frac{m^{n}}{(\gamma_{\mu} D^{\mu})^{n}}
\eq
and only in first term there is factorization, other terms
 may mix  constant and  variable modes. However  their
 contribution is suppressed by powers $\beta m$ and one can
 think about their importance at temperatures $T < m$.
It is still unclear  what is the order parameter
 corresponding to the Hagedorn transition in the theory
 with light fermions with mass $m \leq \sqrt{g^{2}N/\pi}$.
 In any case the usual picture of the $Z_{N}$ symmetry
 breaking is wrong in this case and one has the strange
 picture of unbroken $Z_{N}$ with $<L(x)> = 0$, but
 with nonzero $<L(\infty) L(0)>$  which means deconfining.
\vfill
\eject

\newsection{ Conclusion}

 In this paper we tried to address some at first sight
 different  but, from our point of view,  related
 problems connected with the $Z_{N}$ symmetry in
 the hot gauge theories. We demonstarted that
 allowing  the nontrivial $Z_{N}$ vacua to exist
 one may have  a real problem when studying the
 metastable states which have impossible
 thermodynamic properties. We tried to present
 arguments that such  bubbles cannot exist as
 metastable states at all. However we did not
 explore in this paper another possibility,
 which seems unrealistic, but can not be excluded
 completely. This is the idea that if after all
 our attempts the metastable states will survive
 one must consider them as some kind of states with
 {\bf inverse population}, i.e. the states with
 {\bf negative} temperature!  Let us remember
 that both negative specific heat and negative
 entropy proportional to $T^{3}$ and for negative $T$
 they will change the sign, i.e. in the  case when
 they were negative at positive $T$ they will be
 positive for negative $T$. This idea seems rather
 strange because we still have a paradox - what
 we have started from was the gas of quarks and gluons
 at high positive temperatures. How does one get the
 negative temperature  is a big question. One may speculate that
 we  can reach this region passing through infinte $T$.
 If such a thing would be possible one could imagine
 something like quark-gluon laser in hot gauge theories.
  Despite all it strangeness  this idea deserves further analysis.

We also considered connection between existence of
 $Z_{N}$ bubbles and the  the instabilities
 of the confining phase at high temperatures. Using
 two-dimensional QCD with adjoint fermions as an example
 we had demonstrated that situation with the
 breaking of the $Z_{N}$ symmetry is not so simple
 as one could imagine  when dealing with "simple"
 two-dimensional models. In fact, for massless case
 we have proved that the $Z_{N}$ symmetry can not be broken
 at all, however the correaltion function of the two
 Polyakov lines $<L(x)L(0)>$ is not going to zero when
 $x\ra\infty$ and we are in the deconfining phase at
 high temperatures. It is unclear how the mass of the
 fermion affect this situation. We conjectured that
 the same physics must be correct for light enough
 fermions with mass $m$ less then $\sqrt{g^{2}N}$. For
 heavy fermions with $m^{2}> g^{2}N$ one may have
  another phase  and may be even the breaking of
 global $Z_{N}$. It is amusing that in this theory one has
 hidden supersymmetry precisely at $m^{2} = g^{2}N/\pi$ as was
 shown by Kutasov in \cite{kutasov}. Does it mean that SUSY
 point is a  phase transition point  and one has two phases
 in a  QCD with adjoint fermions is a very intriguing possibility.

It is also unclear how reliable is the effective action in
 the four-dimensional gauge theories at high temperatures.
 What is known is the effective action for the constant
 field and one assumes that it is possible to substitute
 the constant field $A_{0}$ by  general $x$-dependent
 field $A_{0}(x)$.  We just saw, however, that this
 recipe is not good all the time - two-dimensional
 example is a good  demonstartion of this fact. May be
 in four dimensions we also missed something ? In any
 case  it seems that the problem of $Z_{N}$ phases
 and related with it the problem of the instabilities
 of the confining phase \footnote{Which may give us a unique
 information about  structure of QCD strings, if string description
 of QCD exists}  definitely deserve further investigations.

{\bf Acknowledgements.}

\noindent

\bigskip

This results obtained in this paper were presented at the
 ITP at University of California at Santa Barbara
 Workshop "QCD at finite temperature" in August 1993. I would
 like to thank the stuff of ITP for warm hospitality
  and B. Mueller, J.Polchinski, E. Shuryak, J. Verbaarschot and
 especially   I. Zahed
 for interesting discussions during the
  Workshop. It is a pleasure to thank K.Demeterfi,
 D. Gross, I. Klebanov, D. Kutasov,
 A. Polyakov and A. Smilga
  for  stimulating  discussions. This work was supported by the
 National Science Foundation grant NSF PHY90-21984.

{\renewcommand{\Large}{\normalsize}

\end{document}